\documentclass{optica-article}

\journal{opticajournal}
\articletype{Research Article}

\usepackage{lineno}
\begin{document}

\title{Neural-network reconstruction of THz transmission spectra using electrically tunable AlGaN/GaN plasmonic-crystal analyzer}

\author{A. Witkowska,\authormark{1,*} M. Dub,\authormark{1} P. Sai,\authormark{1} P. Tiwari,\authormark{2} M. Sakowicz,\authormark{2} J. A. Majewski,\authormark{1,3} and W. Knap\authormark{1,2}}

\address{\authormark{1}Center for Terahertz Research and Applications -- CENTERA 2, Centre for Advanced Materials and Technologies CEZAMAT, Warsaw University of Technology, ul.~Poleczki 19, 02--822 Warsaw, Poland}
\address{\authormark{2}Institute of High Pressure Physics, Polish Academy of Sciences, ul.~Sokołowska 29/37, 01--142 Warsaw, Poland}
\address{\authormark{3}Faculty of Physics, University of Warsaw, ul.~Pasteura 5, 02--093 Warsaw, Poland}

\email{\authormark{*}agnieszka.witkowska2@pw.edu.pl}

\begin{abstract}
We demonstrate machine learning (ML) based reconstruction of terahertz transmission spectra using an electrically tunable grating-gate AlGaN/GaN plasmonic-crystal analyzer. The analyzer encodes the transmission spectrum into a voltage-dependent intensity, which is then inverted by an ML algorithm. A feedforward neural network trained on a synthetic dataset is validated experimentally on four samples in standard Fourier Transform Infrared (FTIR) mode and in direct (fixed-mirror) acquisition mode. The network achieves a mean square error (MSE) of the reconstruction of $0.015$ in FTIR mode and $0.038$ in direct mode, correctly identifying six out of seven ground-truth resonances in each mode. Against a first-difference Tikhonov regularization baseline, the mean reconstruction error is reduced $3.6$ times in FTIR mode and $1.55$ times in direct mode, with fewer spurious peaks and lower peak-position errors. Voltage-tunable plasmonic filtering combined with neural-network inversion establishes an interferometer-free architecture for THz spectral reconstruction.
\end{abstract}


\section{Introduction}
\label{sec:intro}

Terahertz (THz) spectroscopy is an established tool for
the non-destructive characterization of pharmaceuticals
\cite{king2011diclofenac,strachan2005pharma}, materials
\cite{li2025thickness}, biological tissue
\cite{peng2020cancer,sadeghi2023biomaterials}, and concealed
substances \cite{cheng2025security,zhang2018explosives}. The
principal experimental platforms — time-domain spectroscopy
\cite{neu2018tutorial}, frequency-domain spectroscopy
\cite{lu2023fds}, and Fourier-transform infrared THz spectroscopy
\cite{brundermann2012thztechniques} — achieve high spectral fidelity at
the cost of substantial optical infrastructure: femtosecond lasers,
precision-aligned interferometers, or cryogenically cooled detectors.
Deploying THz spectroscopy beyond the specialist laboratory motivates
architectures that replace spectrometer optics with computation while
retaining spectroscopic selectivity.

Compact, tunable THz filters offer a promising alternative. In this approach, an external control parameter, e.g., gate voltage, varies the transmission of a filter-analyzer, encoding the spectrum of interest into a number of intensity readings, from which it is then reconstructed computationally. The spectrum of interest can be either unknown emission or the transmittance of an unknown sample placed behind the known source and analyzer, as in the present work. Tunable THz filter platforms that have been explored include cyclotron-resonance notch filters \cite{skierbiszewski1998cyclotron}, topological-photonic
notch filters \cite{gupta2023topological}, and waveguide-bandgap
filters \cite{lee2011notch}. Among these, electrically tunable
AlGaN/GaN plasmonic crystals stand out as large-area solid-state
devices with a continuously tunable response: the grating-gated sample
S7 introduced by Sai \emph{et al.}~\cite{sai2023prx} carries
plasmonic resonances that shift continuously with the gate voltage
$V_{\mathrm{G}}$ across the $0.6$--$3.3\,\mathrm{THz}$ band. When placed in
series with an unknown sample of transmittance $T(\nu)$ in front of
a broadband detector, it encodes $T(\nu)$ as a detector signal reading $s(V_G)$, being a function of the applied gate voltage $V_G$
\begin{equation}
s(V_{\mathrm{G}}) \;=\; \int A(\nu, V_{\mathrm{G}})\, T(\nu)\, \mathrm{d}\nu.
\label{eq:forward_intro}
\end{equation}
Here, $A(\nu, V_{\mathrm{G}})$ is the measured forward operator combining the source spectrum, analyzer transmission, and absorption/reflection of the radiation within the experimental setup. Direct inversion of Eq.~(\ref{eq:forward_intro})
is an ill-posed problem for which classical methods—least-squares 
fitting, Tikhonov regularization, and compressive sensing—might 
provide robust mathematical frameworks but suffer from noise
sensitivity, manual parameter tuning, and bias–variance trade-offs~\cite{hansen2010discrete}.

Computational spectral reconstruction methods have been demonstrated with
plasmonic metasurfaces in the mid-infrared using recursive
least-squares inversion~\cite{craig2018metasurface} and with
photonic-crystal slabs in the visible using regularized
optimization~\cite{wang2019singleshot}. Neural-network-based
reconstruction has subsequently been demonstrated on on-chip
plasmonic encoders in the visible and near-infrared
\cite{brown2021plasmonic,zhang2021hyperspectral,tua2023rainbow}.
These platforms are covered in recent reviews of miniaturized
computational spectrometers~\cite{yang2021miniaturization}. In the
THz range, neural-network spectral reconstruction has been
demonstrated theoretically with tunable graphene-plasmonic encoders trained on
full-wave electromagnetic simulations~\cite{he2024tunable}, and with
a magneto-electric metasurface array trained on finite-element
simulations~\cite{zhao2026ultracompact}.

We demonstrate the computational reconstruction of transmission spectra from intensity readings $s(V_G)$ in a system consisting of a broadband source, an analyzer, the investigated sample, and a detector (inside the FTIR measurement setup). We train a neural network model that takes $s(V_G)$ as input and predicts the transmission $T(\nu)$. The model training is performed on a theoretical dataset and is validated on four experimental test samples in two acquisition modes: a standard FTIR configuration and in direct (fixed-mirror) mode.

\section{Experimental setup and measurements}
\label{sec:setup}

Experimental characterization of the analyzer used in this work, as well as measurements in the setup with both the analyzer and test samples, were performed using a Fourier-transform infrared vacuum spectrometer (Bruker Vertex V80v, Billerica, MA, USA) equipped with two precision source/measure units (Keysight Technologies B2902A, Santa Rosa, CA, USA), a continuous-flow liquid-nitrogen cryostat (Oxford Instruments Optistat CF-V, Abingdon, UK), a cryogenically cooled silicon bolometer (Infrared Laboratories General Purpose 4.2 K Bolometer System, Tucson, AZ, USA), a lock-in amplifier (AMETEK 7260, Berwyn, PA, USA), an optical chopper system (Thorlabs MC2000B, Newton, NJ, USA), and a mercury lamp as the radiation source. Temperature stabilization was provided by a temperature controller (Oxford Instruments MERCURY-ITC-1, Abingdon, UK) connected to a heater and a calibrated temperature sensor mounted near the sample on the cold finger inside the cryostat. All measurements were carried out at 70 K. For far-infrared FTIR measurements, a silicon beam splitter was used.

Far-infrared spectra in fast-scan mode were acquired by signal averaging over an integration time of 3 min, with a scanner velocity of 80 kHz and a spectral resolution of 1 cm$^{-1}$. To prevent oversaturation of the Si bolometer, an integrated diamond low-pass filter was employed to suppress radiation above 3.3 THz. Consequently, the transmitted signal was analyzed within the frequency range of 0.6–3.3 THz. On both sides of the cryostat cold finger, 3 mm apertures were positioned to confine the transmitted radiation path, thereby ensuring that the measured transmittance signal originated only from a limited area of the sample. The grating-gate field-effect transistor - analyzer - was mounted on the front aperture, enabling transmission measurements as a function of gate voltage for subsequent analysis. 

In the second stage of the experiment, each sample under test was mounted on the second aperture to investigate the resulting modification of the transmission spectrum. The resulting spectra integrated over frequency are referenced as the FTIR mode measurements $s_\mathrm{{FTIR}}(V_G)$.

In the third stage of the study, the detector response was measured as a function of the gate voltage applied to the analyzer. For these measurements, the spectrometer mirrors were fixed at the zero optical path difference (ZPD) position, while signal modulation was provided by an optical chopper integrated into the spectrometer chamber. The detector output was demodulated using a lock-in amplifier and digitally recorded, yielding the signal amplitude and phase as functions of gate voltage. During the experiment, the gate voltage was controlled using a precision source$/$measure unit in the range from 0 to -10 V with a step of 0.01 V. The optical chopper operated at 383 Hz, and the lock-in integration time was set to 200 ms. Each data point was acquired after a 1 s voltage stabilization period. As a reference, the signal transmitted through the grating-gate field-effect transistor and the empty second aperture was measured first. Subsequently, each sample under the test was mounted on the second aperture to investigate the resulting modification of the transmission spectrum. The resulting signals are referenced as the direct mode measurements $s_{\mathrm{dir}}(V_G)$ .

\subsection{The plasmonic-crystal analyzer}
\label{sec:S7}
\begin{figure}[htb]
\centering
\includegraphics[width=0.98\linewidth]{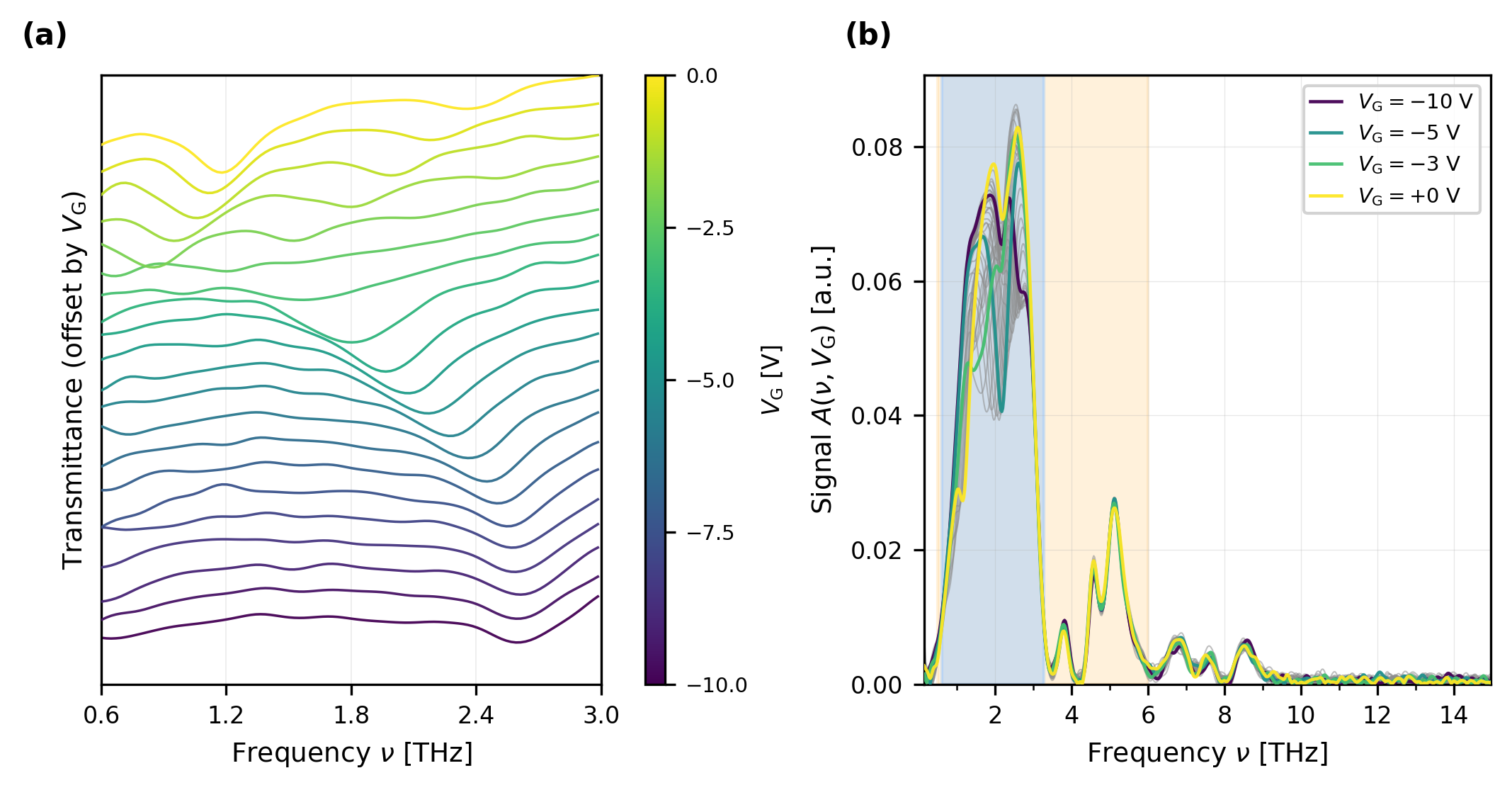}
 \caption{(a) FTIR measured transmittance of the S7 analyzer for various applied gate voltages, in the frequency range where plasmonic resonances occur; spectra are plotted with a vertical offset for clarity, (b) FTIR signal measured in the configuration with the analyzer alone, constituting the analyzer response matrix $A(\nu, V_G)$. Gray lines correspond to the signal for all $V_G$ values, while coloured curves highlight representative biases. The orange band marks the integration window of Eq.~\ref{eq:forward_intro} used for synthetic dataset generation, and the blue band indicates the frequency range over which the ML model performs the reconstruction.}
    \label{fig:analyzer}
\end{figure}

The analyzer used throughout this work is the plasmonic crystal S7, described in detail in Ref.~\cite{sai2023prx}. It is an AlGaN/GaN quantum well deposited on a $500\,\mu\mathrm{m}$ semi-insulating SiC substrate, with a two-dimensional electron gas (2DEG) at the AlGaN/GaN interface and a metallic grating gate of period $a_{\mathrm{G}} = 1.5\,\mu\mathrm{m}$ and filling factor
$f = 0.6$ patterned over a $1.7\times 1.7\,\mathrm{mm^{2}}$ active
area.
The transmisttance of the analylzer is varied with variying voltage, which is shown in  Fig.~\ref{fig:analyzer}(a). FTIR signal recorded in the setup with analyzer only is denoted $A(\nu, V_{\mathrm{G}})$ and it is shown in Fig.~\ref{fig:analyzer}(b). It factorizes as the product of analyzer transmittance (dependent on the $V_G$), the Hg-arc source spectrum, and the $V_{\mathrm{G}}$-invariant FTIR instrument response (beamsplitter, cryostat windows, bolometer responsivity, low-pass filter).

\subsection{Test samples}
\label{sec:samples}

To experimentally validate the proposed computational reconstruction methodology, we selected four samples exhibiting absorption of THz radiation within the active band of our analyzer. The samples, together with the positions of their transmission minima, are listed as follows: ARef at 0.94 and 2.48 THz, CRef at 1.03 and 2.20 THz, at 2.62 THz, and F06 at 1.36 and 2.39 THz. Both the signals acquired in the two measurement modes and the full transmission spectrum are presented in Fig.~\ref{fig:real_ftir}.

\section{Computational Reconstruction Methods}
\label{sec:methods}

\subsection{Dataset generation}
\label{sec:db8}
\label{sec:forward}

We generated a dataset for training the neural network (NN) model, consisting of paired data: the detector signal as a function of gate voltage $V_{\mathrm{G}}$ (input to the model) together with the transmittance spectrum $T(\nu)$ that produces this signal. Ideally, one would rely exclusively on experimental data; however, due to the limited availability of test samples, we used a set of $N \approx 1.4 \cdot 10^{5}$ theoretically generated transmittance spectra, a quantity that would be impossible to obtain experimentally within a reasonable time.

To represent the spectra of interest, we generated transmission profiles with zero to five minima, each described by a Lorentzian or Gaussian line shape parametrized by centre position, FWHM, and amplitude, and modulated by a baseline accounting for possible substrate responses. Peak counts $n_p \in \{0,\dots,5\}$ were stratified with the majority weight placed on the $1$--$3$-peak regime, consistent with the test samples discussed in Sec.~\ref{sec:samples}. Within each stratum, parameters were drawn from a quasi-random Sobol sequence.

For each theoretical transmittance spectrum the detector signal was generated by evaluating Eq.~\ref{eq:forward_intro} numerically over the integration band $\nu \in [0.5,\,6]\,\mathrm{THz}$. This range brackets the spectral interval in which the signal from the mercury lamp, transmitted through the analyzer and measured by the bolometer, is non-negligible, as shown in Fig.~\ref{fig:analyzer}(b). The transmittance was sampled on a uniform grid of $117$ points $\{\nu_j\}_{j=1}^{117}$ spanning $[0.6,\,3.3]\,\mathrm{THz}$ -- the tunability range of the plasmonic resonances of the analyzer -- while the detector signal was sampled on the experimental grid of $41$ gate-voltage values $\{V_{\mathrm{G},i}\}_{i=1}^{41}$. Each spectrum was further augmented with four Gaussian noise realisations to train the network for robustness against measurement noise, altogether yielding more than $5\cdot 10^{5}$ paired vectors $(\mathbf{s},\,\mathbf{T})$, where $\mathbf{s}\in\mathbb{R}^{41}$ collects the detector signals $s(V_{\mathrm{G},i})$ and $\mathbf{T}\in\mathbb{R}^{117}$ samples the transmittance $T(\nu_j)$.

\subsection{Network architecture and training}
\label{sec:nn}
\label{sec:split}

The full set of transmission spectra was split $70\,\% / 15\,\% / 15\,\%$ into training, validation, and test sets, stratified by peak count $n_p$, with all noise realizations of a given spectrum constrained to the same split so that no spectrum leaks across sets.

For the reconstruction network we chose a fully connected neural network $f_{\boldsymbol{\theta}}:\mathbb{R}^{41}\!\to\!\mathbb{R}^{117}$, mapping the detector-signal vector $\mathbf{s}$ onto the reconstructed transmittance $\hat{\mathbf{T}}_{\mathrm{NN}} = f_{\boldsymbol{\theta}}(\mathbf{s})$, sampled on the same frequency grid $\{\nu_j\}_{j=1}^{117}$. The model consists of six hidden layers of $256$ units each, with every layer followed by batch normalization and a Swish activation~\cite{ramachandran2017swish}, and dropout applied at decreasing rates across the first five blocks. The output layer, with sigmoid activation, enforces $\hat{T}_{\mathrm{NN}}(\nu_j)\in(0,1)$ for every $j$. The network has $\sim 3.8\times 10^{5}$ trainable parameters $\boldsymbol{\theta}$. Training employed the AdamW optimizer~\cite{loshchilov2019decoupled} (learning rate $10^{-3}$, weight decay $10^{-4}$) with the loss
\begin{equation}
    \mathcal{L}(\boldsymbol{\theta}) \;=\; \bigl\|\,\hat{\mathbf{T}}_{\mathrm{NN}} - \mathbf{T}\,\bigr\|_{2}^{2},
    \label{eq:nn_loss}
\end{equation}
batch size $128$, early stopping on validation loss, and a plateau-triggered learning-rate schedule. This configuration was selected as the winner of a 34-configuration hyperparameter grid search over layer stack, weight decay, batch size, loss type, learning rate, and dropout schedule. Five independent seeds were subsequently trained with these winning hyperparameters; all fall within $0.6\,\%$ of the best validation MSE, indicating that seed variance is negligible. The seed with the lowest validation loss was used as the production model for all subsequent evaluations.

\subsection{First-difference Tikhonov regularization -- a classical baseline}
\label{sec:tikhonov}

To benchmark the proposed NN model against an established classical method, we implement first-difference Tikhonov regularization~\cite{hansen2010discrete}, which seeks the solution of the inverse problem defined by Eq.~\ref{eq:forward_intro}. Discretized on the grids $\{V_{\mathrm{G},i}\}$ and $\{\nu_j\}$, the forward model takes the matrix form
\begin{equation}
    \mathbf{s} \;=\; \mathbf{A}\,\mathbf{T},
    \label{eq:forward_discrete}
\end{equation}
where $\mathbf{A}$ is the discretized analyzer response (Fig.~\ref{fig:analyzer}), and $\mathbf{s}$, $\mathbf{T}$ are the signal and transmittance vectors introduced in Sec.~\ref{sec:forward}. The Tikhonov-regularized estimate is then obtained as
\begin{equation}
    \hat{\mathbf{T}}_{\mathrm{Tikh}} \;=\; \arg\min_{\mathbf{T}}\,\bigl\| \mathbf{s} - \mathbf{A}\,\mathbf{T} \bigr\|_{2}^{2} \;+\; \alpha\,\bigl\|\mathbf{D}_{1}\,\mathbf{T}\bigr\|_{2}^{2},
    \label{eq:tikhonov}
\end{equation}
where $\mathbf{D}_{1}$ is the first-difference operator that penalises non-smooth solutions. The regularization strength $\alpha$ is selected per sample by generalised cross-validation (GCV).

\section{Results}
\label{sec:results}
\subsection{Synthetic test-set performance}
\label{sec:r_synth}

On the held-out synthetic test set, the trained model achieves a mean per-spectrum mean square error (MSE) of $5.7\times 10^{-3}$ and a pooled test $R^{2} = 0.934$. Per-class reconstruction error grows monotonically with the number of peaks $n_{\mathrm{p}}$: $\mathrm{MSE} = 1.0\times 10^{-3}$ at $n_{\mathrm{p}}=0$ (no peaks, near-flat targets), $4.0\times 10^{-3}$ at $n_{\mathrm{p}}=1$, and saturating around $\sim\!8\times 10^{-3}$ for $n_{\mathrm{p}}\geq 3$.

\begin{figure}[htb]
\centering
\includegraphics[width=\linewidth]{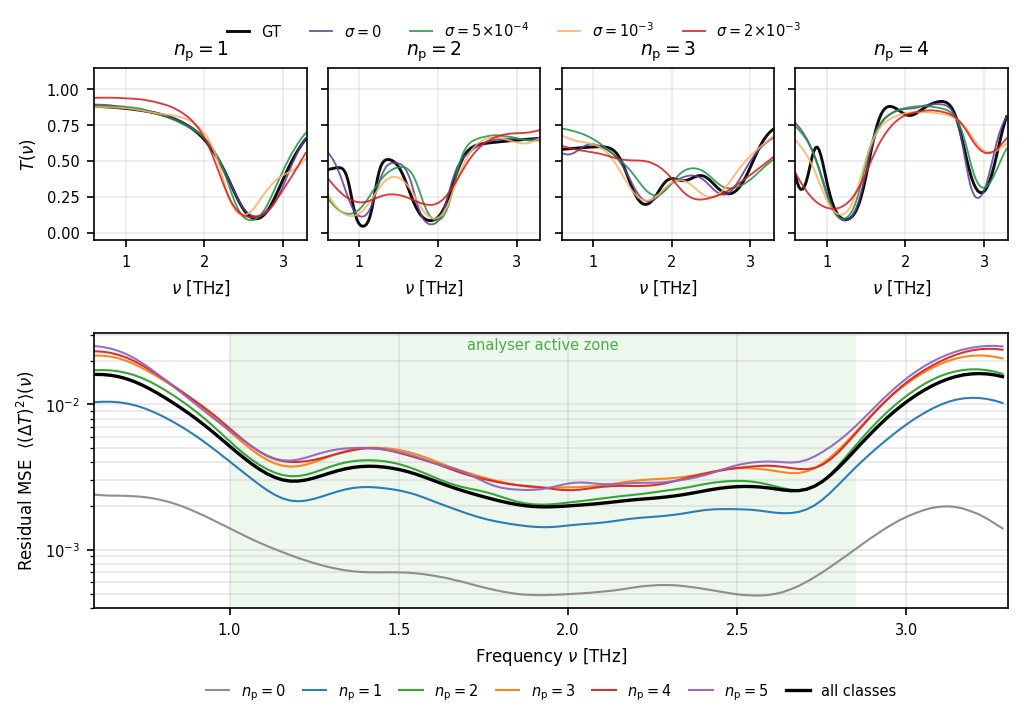}
\caption{Synthetic test-set reconstruction.
Top row: one random example per generator-level peak-count class
$n_{\mathrm{p}} = 1$--$4$, showing the ground-truth $T(\nu)$ (black, thick) and the NN reconstruction $\hat{T}_{\mathrm{NN}}(\nu)$ at the four trained noise levels ($\sigma = 0$,
$5\times 10^{-4}$, $10^{-3}$, $2\times 10^{-3}$).
Bottom: frequency-resolved residual MSE
$\langle (T - \hat{T}_{\mathrm{NN}})^{2}\rangle(\nu)$
averaged over all test set, with per-$n_{\mathrm{p}}$
decomposition. $1.0$--$2.85\,\mathrm{THz}$ area shaded, where MSE is lowest.}
\label{fig:synth_test}
\end{figure}

Fig.~\ref{fig:synth_test} (top) shows reconstructions at the four trained noise levels $\sigma$ for exemplary synthetic spectra. Single-peak ($n_{\mathrm{p}}=1$) and paired-peak ($n_{\mathrm{p}}=2$) ground-truth (GT) spectra $T(\nu)$ are recovered with high fidelity, with the reconstructions $\hat{T}_{\mathrm{NN}}(\nu)$ at $\sigma = 0$, $5\times 10^{-4}$ and $10^{-3}$ overlapping closely and only $\sigma = 2\times 10^{-3}$ producing visible residual deviation. For $n_{\mathrm{p}} \geq 3$ the reconstructions track the dominant resonance structure but tend to merge or omit the finer features that overlap into a few visible dips. The frequency-resolved residual MSE $\langle (T - \hat{T}_{\mathrm{NN}})^{2}\rangle(\nu)$ (Fig.~\ref{fig:synth_test}, bottom) is lowest in the $1.0$--$2.85\,\mathrm{THz}$ active zone and rises by a factor of $\sim\!4$ at the band edges. This bandwidth-limited conditioning is a property of the analyzer's reduced $V_{\mathrm{G}}$ modulation at the band edges (Fig.~\ref{fig:analyzer}).

\subsection{Real samples: FTIR mode}
\label{sec:r_ftir}
In order to evaluate the model on experimental data, we use integrated FTIR measurements $s_{\mathrm{FTIR}}(V_{\mathrm{G}})$ of the radiation from the source passing through the analyzer and test samples (as described in Sec.~\ref{sec:setup}. $s_{\mathrm{FTIR}}(V_{\mathrm{G}})$ is the input to the model, and the reconstructed transmittance $\hat{T}_{\mathrm{NN}}(\nu)$ is compared to the ground-truth transmittance $T(\nu)$.

\begin{figure}[htb]
\centering
\includegraphics[width=1\linewidth]{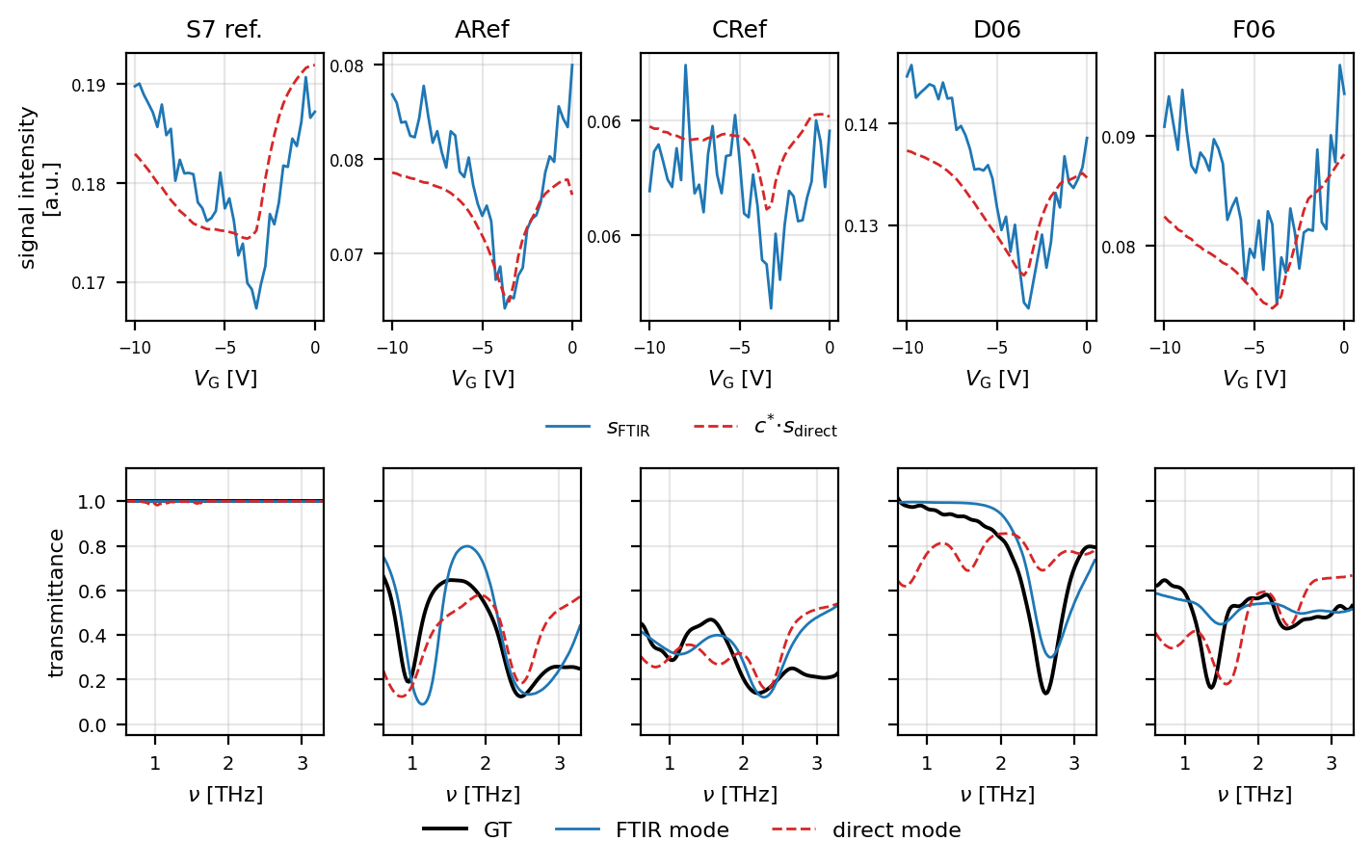}
\caption{Comparison of the signal in FTIR and direct acquisition modes for the reference (no sample) and four test samples with reconstruction in both modes. (Top row) $s_{\mathrm{FTIR}}(V_{\mathrm{G}})$ (blue solid) and
$c^{*}\!\cdot\!s_{\mathrm{direct}}(V_{\mathrm{G}})$ (red dashed),
with $c^{*}=6.09$ (details in the text). (Bottom row)
NN reconstruction $\hat{T}_{\mathrm{NN}}(\nu)$ in FTIR mode (blue solid) and
 direct mode (red dashed) against the ground-truth $T(\nu)$ (black).}
\label{fig:real_ftir}
\end{figure}

The mean per-sample MSE is $0.015$. The NN model locates six of seven ground-truth resonances with zero spurious detections. The mean peak-position MAE on matched peaks is $3.7\,\mathrm{cm^{-1}}$, with per-peak errors spanning $2.3$--$6.2\,\mathrm{cm^{-1}}$. Peak transmittance minima $T_{\min}$ are recovered with mixed direction of error and $|\Delta T_{\min}|\!\le\!0.16$ absolute across five of the six matched peaks; F06 at $1.36\,\mathrm{THz}$ is overestimated by $0.29$ ($T_{\min}\!\approx\!0.16\rightarrow 0.45$). The deepest ground-truth resonance (D06, $T_{\min}\!\approx\!0.14$) is reconstructed at $T_{\min}\!\approx\!0.30$, an underestimate of $0.16$. The detailed summary is presented in Tab.~\ref{tab:ftir_direct}.

\begin{table}[htb]
\centering
\caption{Per-sample reconstruction error and peak-position MAE for the NN model in FTIR and direct modes. Here we denote match/GT: matched predicted peaks (prominence 0.05, 0.4\,THz tolerance) against GT count; unmatched predictions are spurious. The S7 reference (flat $T\!\equiv\!1$) is included for completeness.}
\label{tab:ftir_direct}
\setlength{\tabcolsep}{4pt}
\small
\begin{tabular}{lcccc|cccc}
\hline
 & \multicolumn{4}{c|}{FTIR mode} & \multicolumn{4}{c}{Direct mode} \\
sample & MSE & match/GT & spur. & MAE\,[cm$^{-1}$] & MSE & match/GT & spur. & MAE\,[cm$^{-1}$] \\
\hline
S7 ref. & $<\!10^{-5}$ & \textendash & 0 & \textendash & $2\!\times\!10^{-5}$ & \textendash & 0 & \textendash \\
ARef & 0.025 & 2/2 & 0 & 5.0 & 0.035 & 2/2 & 0 & 1.5 \\
CRef & 0.017 & 2/2 & 0 & 3.5 & 0.027 & 1/2 & 0 & 3.9 \\
D06  & 0.011 & 1/1 & 0 & 2.3 & 0.064 & 1/1 & 1 & 1.5 \\
F06  & 0.009 & 1/2 & 0 & 3.9 & 0.026 & 2/2 & 1 & 4.2 \\
\hline
summary & 0.015 & 6/7 & 0 & 3.7 & 0.038 & 6/7 & 2 & 2.8 \\
\hline
\end{tabular}
\end{table}

\subsection{Real samples: direct acquisition mode}
\label{sec:r_direct}

In direct mode (described in Sec.~\ref{sec:setup}), the mirrors are fixed in the ZPD position. Both modes share the same bolometer detector but differ in signal modulation: FTIR mode acquires interferograms by scanning the moving mirror, while direct mode chops the Hg-arc source at $383\,\mathrm{Hz}$ and recovers a single bolometer intensity per $V_{\mathrm{G}}$ via lock-in demodulation. This results in varied shape and amplitude of the recorded signal, as shown in Fig.~\ref{fig:real_ftir} (top row). The differences arise likely from the bolometer's frequency-dependent responsivity and the different signal-processing chains, but the precise contributions are not separately quantified.
The amplitude of the signal in direct mode is corrected by a single global rescaling factor $c^{*}=6.09$, calibrated on the reference measurement (without the test sample) by least-squares fit over the 41 voltage values. The factor agrees within $0.4\,\%$ with the $L_{2}$-norm and integral ratios, is seed-independent, and is not tuned per sample.

The mean per-sample MSE in direct mode is $0.038$, and is higher compared to the FTIR-mode value of $0.015$. The pooled $R^{2}$ is negative on all four samples, reflecting residual differences between the rescaled direct signal and the training distribution (visible in Fig.~\ref{fig:real_ftir}, top row). The peak-position metric nevertheless survives this distribution shift: six of seven resonances are recovered at $2.8\,\mathrm{cm^{-1}}$ mean MAE, with per-peak errors spanning $0$--$6.9\,\mathrm{cm^{-1}}$. The missed CRef $1.02\,\mathrm{THz}$ feature and the two spurious detections on D06 and F06 show limited ability of the model to recover spectra from direct measurement mode. Peak depth recovery degrades on the deepest resonance: D06's GT $T_{\min}\!\approx\!0.14$ is reconstructed at $T_{\min}\!\approx\!0.69$ (an underestimate of $0.55$), while the other matched peaks remain within $|\Delta T_{\min}|\!\le\!0.07$.

\subsection{Benchmark against first-difference Tikhonov regularization}
\label{sec:r_tikhonov}

\begin{figure}[htb]
\centering
\includegraphics[width=\linewidth]{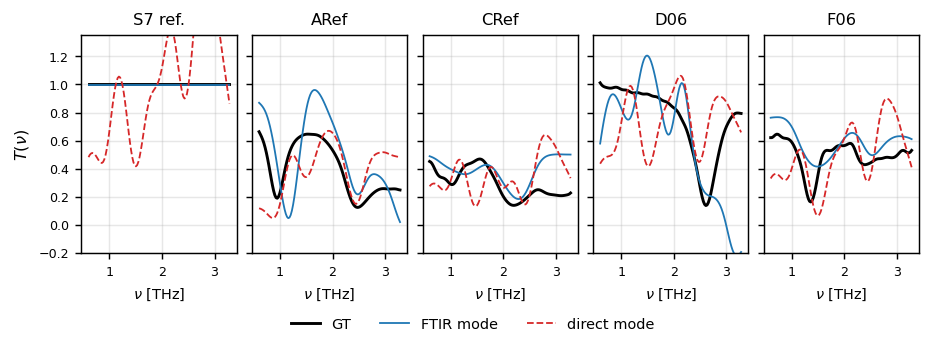}
\caption{First-difference Tikhonov reconstruction $\hat{T}_{\mathrm{Tikh}}(\nu)$ under GCV-selected $\alpha$ for the reference (no sample, $T\!\equiv\!1$) and four test samples in FTIR (solid blue) and direct (dashed red) modes, compared against the ground-truth transmittance $T(\nu)$ (black).}
\label{fig:tikhonov}
\end{figure}

To place the neural-network performance in the context of classical inverse methods, we solve the same reconstruction problem with first-difference Tikhonov regularization (Eq.~\ref{eq:tikhonov}) on identical inputs, with the regularization parameter $\alpha$ selected per sample by GCV (Fig.~\ref{fig:tikhonov}).

In FTIR mode the NN model reduces mean MSE by 3.6 times relative to Tikhonov ($0.015$ vs.\ $0.055$). In direct mode the advantage narrows to 1.55 times ($0.038$ vs.\ $0.059$). Under matched detection criteria, the NN model yields fewer spurious detections (0 vs.\ 2 in FTIR, 2 vs.\ 3 in direct) and lower peak-position MAE ($3.7$ vs.\ $4.6\,\mathrm{cm^{-1}}$ in FTIR, $2.8$ vs.\ $4.2\,\mathrm{cm^{-1}}$ in direct). The neural network, therefore, produces reconstructions both more accurate in mean-square sense and cleaner in resonance-position content.

\section{Conclusion}
\label{sec:conclusion}
We demonstrate neural network reconstruction of THz transmission spectra using single electrically tunable grating-gate AlGaN/GaN plasmonic crystal as analyzer. The analyzer encodes the transmission spectrum into a voltage-dependent intensity reading. Training of the NN model relies entirely on a synthetic dataset, with experimental validation performed against measured transmission spectra in two acquisition modes.

In FTIR mode the model reaches a mean MSE of $0.015$ and peak-position MAE of $3.7~\mathrm{cm}^{-1}$ whereas in direct mode the model reaches MSE $0.038$ and MAE $2.8~\mathrm{cm}^{-1}$ after a single global rescaling factor calibrated on a sample-free reference. Six of seven ground-truth resonances are correctly identified in each mode. Against first-difference Tikhonov regularization on identical inputs, the network reduces mean MSE by $3.6$ and $1.55$ times in FTIR in direct modes, respectively, with fewer spurious peaks and lower peak-position errors.

Two aspects shape the present implementation. The deepest absorption resonances are systematically underestimated in depth, likely reflecting the bias of a model trained on synthetic spectra toward the depth distribution of its training set. Reconstruction fidelity also drops at the edges of the $1.0$--$2.85~\mathrm{THz}$ active zone, narrower than the analyzer's nominal $0.6$--$3.3~\mathrm{THz}$ tuning range, where the $V_\mathrm{G}$-induced spectral modulation weakens. These caveats aside, voltage-tunable plasmonic filtering combined with neural network inversion establishes an interferometer-free architecture for THz spectral reconstruction.


\begin{backmatter}
\bmsection{Funding}
This work was supported by “Center for Terahertz
Research and Applications (CENTERA2)” project (FENG.02.01-IP.05-T004/23) carried out
within the “International Research Agendas” program of the Foundation for Polish Science, cofinanced by the European Union under European Funds for a Smart Economy Programme.

This work was partially funded by National Science Centre, Poland, grant number 2020/38/E/ST7/00476.

\bmsection{Disclosures}
The authors declare no conflicts of interest.

\bmsection{Data availability}
The code, trained model weights, and evaluation data underlying the results presented in this paper will be made available by the authors on reasonable request. Please contact the corresponding author at agnieszka.witkowska2@pw.edu.pl
\end{backmatter}

\bibliography{refs}

\end{document}